# Controlling ultrafast laser writing in silica glass by pulse temporal contrast

Yuhao Lei,[1,a)] Huijun Wang,[1] Gholamreza Shayeganrad,[1] Yuri Svirko,[2] and Peter G. Kazansky[1,a)]

[1]*Optoelectronics Research Centre, University of Southampton, Southampton, SO17 1BJ, United Kingdom*

[2]*Center for Photonics Sciences, University of Eastern Finland, FI-80100 Joensuu, Finland*

[a)] **Authors to whom correspondence should be addressed:** *yuhao.lei@soton.ac.uk* and *pgk@soton.ac.uk*

**Abstract:** Control of laser writing in transparent dielectrics using temporal contrast of light pulses is demonstrated. Anisotropic nanopores in silica glass are produced by high-contrast of $10^7$ femtosecond Yb: KGW laser pulses rather than low-contrast of $10^3$ Yb fiber laser pulses. Low-contrast pulses are useful for creating lamellar birefringent structures, possibly driven by quadrupole nonlinear current. The difference originates in the fiber laser storing third of its energy in a post-pulse of 200 ps duration. The absorption of this low-intensity fraction of the pulse by laser-induced transient defects with relatively long lifetime and low excitation energy, such as self-trapped holes, drastically changes the kinetics of energy deposition and type of material modification.

The interaction of intense ultrashort light pulses with transparent materials has attracted considerable interest due to its strong application potential. Since the first demonstration three decades ago,[1] the femtosecond micromachining had found a wide range of applications spanning from eye surgery [2] to fabrication of photonic components [3-5] and optical data storage.[6] Rapid deposition of energy carried by sub-picosecond light pulses into transparent dielectric involves a complex physical phenomena,[7-10] driven by the generation of free electrons in the irradiated volume via multiphoton, tunneling and avalanche ionization.[11] The free electron may also occur due to multiphoton ionization of self-trapped excitons and other defects, which are formed within the silica glass bandgap[12] and can be detected by using fluorescence, electron spin resonance, and other techniques.[13,14] Inverse Bremsstrahlung absorption in free electron ensemble leads to irreversible permanent modification of the material[2] due to the energy transfer to the lattice within the electron-phonon relaxation time.[15]

Depending on the conditions, permanent modification of the silica glass under irradiation with femtosecond optical pulses manifests itself as the refractive index increase (often referred as type 1 modification),[3] form birefringence created subwavelength spaces nanoplatelets or nanogratings (type 2),[16] flattened nanopores (type X) [17] and voids (type 3).[6] Nanostructured silica glass has been used for selective etching,[4] polarization beam shaping [5] and five-dimensional (5D) optical data storage.[18]

However, it is worth noting that interpretation and explanation of the physical mechanisms underlying the observed phenomena is rather difficult task because the resulting modification of silica glass may depend not only on the peak intensity of the laser pulses, but also on the type of laser used in the experiment. Specifically, at the same peak intensity, the femtosecond pulses generated by fiber lasers increase the refraction index of silicon in the irradiated volume,[19,20] while pulses from an optical parametric amplifier (OPA) pumped by a Ti: Sapphire laser[21,22] does not. Such a difference has been explained by the fact that the temporal contrasts of light pulses produced by these laser systems [23] are different. Specifically, a fiber laser typically generates femtosecond pulses having sub-nanosecond pedestal comparable with lifetime of free electrons in silicon.[24] This pedestal reduces the temporal contrast of the laser pulses and can essentially influence the modification. However, this phenomenon was ignored in ultrafast laser writing experiments in transparent dielectrics, possibly because it was believed that the low-intensity pedestal could not be absorbed by wide-gap materials.

Here we observed that the sub-nanosecond pedestal can also drastically affect the writing ability of a femtosecond laser pulses in silica glass. We show that a fiber laser operating at sub-MHz repetition rate is not suitable for formation of birefringent nanopore structures (type X modification) in silica glass, however it can be used to produce highly birefringent lamella-shaped voxels (type 2m modification). In contrast, the Yb: KGW laser can write anisotropic structures with flattened nanopores (type X) or isotropic modifications with an increased refractive index (type 1) in a wider range of parameters. The pulse contrast measurements reveal correlation between the micromachining performance and presence of the subnanosecond pedestal in the femtosecond laser pulse. Our analysis shows that the energy stored in the pulse pedestal is also uploaded into the irradiated volume despite its low intensity. This process is governed by light absorption by laser-induced defects created by femtosecond pulse. These defects such asself-trapped holes (STHs) possess a low excitation potential enabling deposition of the pedestal energy into irradiated volume and, correspondingly, affecting the writing performance.

We compare the writing performance of two laser systems operating the same wavelength of 1030 nm The first one is Pharos (Light Conversion Ltd) based on the mode-locked regeneratively amplified Yb-doped potassium gadolinium tungstate (Yb: KGW) laser. The second system is Satsuma (Amplitude Co), which is based on a Yb-doped fiber laser. The Satsuma beam was expanded by a telescope system to match the Yb:KGW beam diameter. For both systems pulse duration $\tau$, repetition rate RR and pulse energy $E_p$, were the same and were varied in the range of 270-500 fs, 100 kHz-1 MHz and 450 nJ -1μJ, respectively.

The temporal contrast of laser pulse was measured by a high dynamic range autocorrelator (pulseCheck SM Type 2, APE). The laser beam was focused via a 0.16 or 0.4 NA aspheric lens 170 μm below the surface of a synthetic silica glass substrate, which was mounted on an XYZ linear air-bearing translation stage (Aerotech Ltd.). The retardance and slow axis azimuth of laser-induced modifications were analyzed with an Olympus BX51 optical microscope equipped with a birefringence measurement system (CRi Abrio imaging system) operating at 546 nm wavelength. The refractive index changes were characterized with a wavefront sensor (SID4-HR, Phasics) mounted on the same microscope.

Both lasers were used to imprint voxels at RR = 500 kHz, $\tau$ = 300 fs, $E_p$ =700 nJ, NA = 0.16, and scan speed of 1 mm/s. Figure 1 show the visualized birefringence written in silica with different number of pulses ($N_p$) using both laser systems. One can observe from Fig. 1a that when writing with the Yb:KGW laser, no noticeable induced birefringence was observed at $N_p$ < 20, while at 20 < $N_p$< 150, type X birefringent voxels, which are not seen in transmittance, were observed. At $N_p$ > 150, birefringent voxels begin to appear in transmission, which indicates an increase in the size and their clustering, and randomly distributed nanolamellae formed (type X- 2). When the number of pulses exceeds 400, lamella-like birefringent structures oriented perpendicular to the plane of polarization of the laser beam appears on birefringent and optical images, associated with type 2m modification. When the laser pulse energy was increased up to 800 nJ, at $N_p > 500$, we observed periodic visible lamella-shaped structures (type 2m) separated by $\lambda/n \approx$ 700-800 nm, where $\lambda$ = 1030 nm is the excitation wavelength and n =1.45 is the silica refractive index. It is worth noting that such lamella-shaped structures were with period of $\lambda/n$ were previously observed only in SEM images of imprinted modifications.[25,26] A table was used to compare laser-induced different birefringent modifications in silica glass, as shown in Table S1.

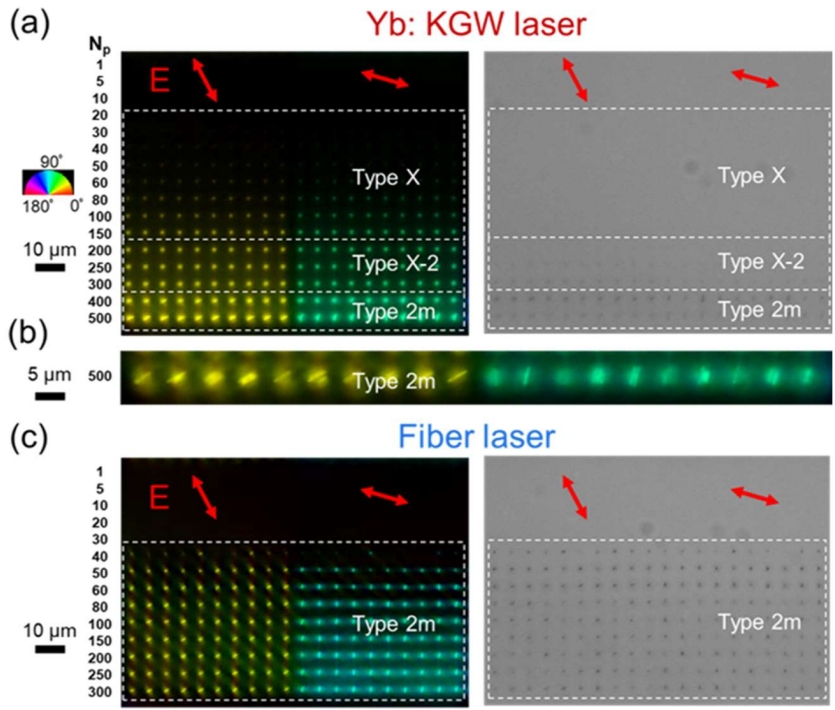

FIG. 1. Birefringent voxels written by (a, b) Yb: KGW laser, and (c) fiber laser. Birefringence image (left) and corresponding optical transmission image (right) of imprinted voxels were captured. The arrow shows the polarization of the laser beam and pseudo-colors (inset) indicate the local orientation of the slow axis. Processing conditions for (a) and (c): 1030 nm wavelength, 300 fs pulse duration, 700 nJ pulse energy, 500 kHz repetition rate, 1-500 number of pulses (Np), focusing via 0.16 NA lens. Ep=800 nJ in (b).

Surprisingly, no "optically invisible" birefringent voxels (type X) were written by using the fiber laser with the same parameters. One can observe from Fig 1c that no modification has been observed at $N_p < 40$, while at $40 < N_p < 300$, birefringent lamella-shaped voxels (type 2m), which are seen both in the birefringence and optical transmission, were imprinted.

When writing was performed with the Yb:KGW laser, the slow axis of birefringent voxels (type X at $N_p$=50, type X-2 at $N_p$=200 and type 2m at $N_p$=500) was always perpendicular to the polarization azimuth of the writing laser beam [Fig. 2a]. Figure 2b shows modification along the beam propagation axis where the writing laser beam propagates from the top to the bottom in the image. The 57 μm long modified area is uniform and comprises randomly arranged anisotropic nanopores (type

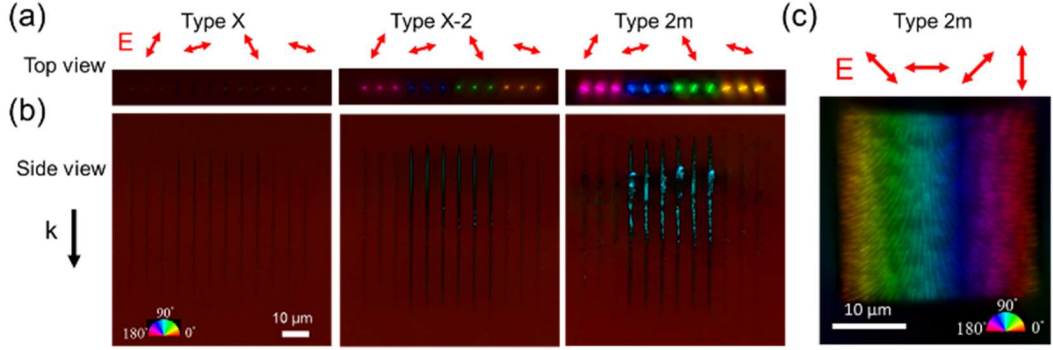

FIG. 2. Birefringence image of different types of modifications written by the Yb: KGW laser. (a) Top view and (b) side view of type X (left), type X-2 (middle) and type 2m (right) voxels. The Np is 60, 200 and 500 for type X, type X-2 and type 2m, respectively. The $E_P$ is 800 nJ and the RR is 500 kHz. (c) Top view of lamella based square (type 2m) written by raster scanning with line separation of 1 μm, Ep of 500 nJ and RR of 200 kHz. E and k: electric field and wave vector of the writing laser beam. Pseudo-colors (inset) indicate azimuth of the slow axis. Processing conditions: 1030 nm wavelength, 300 fs pulse duration, 1 mm/s scan speed, 0.16 NA lens.

X) [Fig. 2b, left]. On the contrary, at $N_P$=500, type X-2 modification with a length of 65 μm comprises birefringent dot-shaped structures in the middle of the modified area [Fig. 2b, middle]. At $N_P$=500, lamella-shaped type 2m modifications are observed at the upper part of the modified area [Fig. 2b, right]. Moreover, lamellae with the wavelength periodicity oriented perpendicular to the polarization direction of writing laser beam were also observed in the structures written by raster scanning with 1 μm line separation, scanning speed of 1 mm/s and pulse energy of 500 nJ [Fig. 2c].

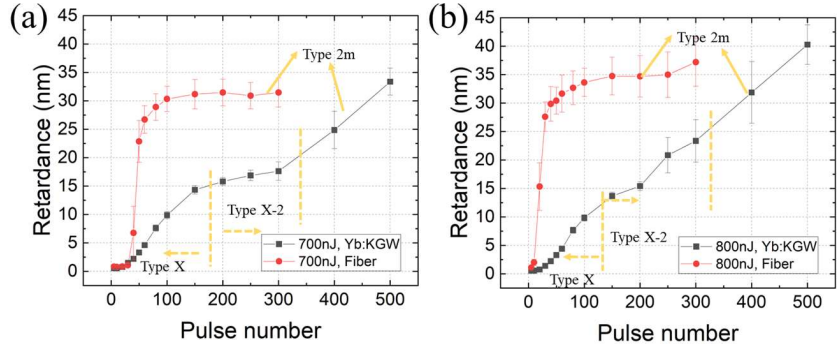

FIG. 3. Retardance of birefringent voxels versus writing pulse number for two lasers. (a) Pulse energy of 700 nJ and (b) 800 nJ. Processing conditions: λ=1030 nm, τ =300 fs, RR=500 kHz, 0.16 NA lens.

Figure 3 shows the retardance of birefringent voxels written by using Yb:KGW and fiber lasers. At the pulse energy of 700 nJ, Yb:KGW laser pulses produced type X voxels at $N_P$≤150 and type X-2 voxels at 200 <$N_P$<300 [Fig. 3a]. The later finding is consistent with previously reported results.[27] In contrast, Satsuma laser produces type 2m modification (i.e., lamella-shaped structures), in which retardance rapidly increases with the number of pulses and saturates at $N_P$ ≈100. At $N_P$ = 200, the retardance $Ret = \Delta n_b l$, where $\Delta n_b$ and $l$ are birefringence and length of the modified area of type 2m voxels written by the fiber laser is about two times higher than that of type X-2 voxels imprinted by the Yb:KGW laser with the same parameters. The retardance of type 2m voxels written with Yb:KGW or fiber lasers are similar, which is about 33 (31) nm for 500 (300) pulses from the Yb:KGW (fiber) laser at $E_P$=700 nJ. Given by voxel length $l$ = 65 μm, one can estimate the birefringence for voxels written with 200 pulses of the fiber and Yb:KGW lasers are 4.9 × 10$^{-4}$ and 2.3 × 10$^{-4}$, respectively. A similar trend was observed at writing pulse energy of 800 nJ [Fig. 3b].

To further elucidate the difference in the parameter windows for obtaining the type X modification, experiments were carried out using different repetition rates, pulse numbers, energies, and durations for both lasers. While the type X modification was produced with Yb:KGW laser at the pulse duration of 300 fs and a repetition rate of 500 kHz in the pulse energy range from 450 nJ to 900 nJ, no modification was produced by the fiber laser [Fig. 4a] was unable to produce any detectible modifications at a pulse energy of less than 700 nJ at the same repetition rate. That is one can conclude that since the modification threshold for the fiber laser is about 1.5 times larger than that of the Yb:KGW laser, one may expect that the actual fiber laser intensity is lower than that of the Yb:KGW laser. However, at the pulse energy above 700 nJ, the fiber laser pulses produces lamella-shaped voxels having stronger birefringence in comparison with obtained by the Yb:KGW laser.

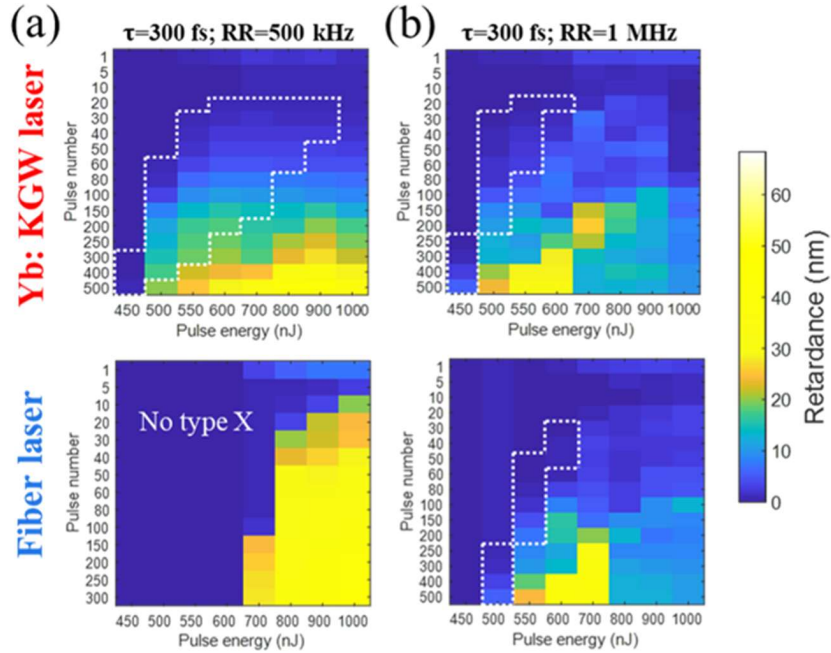

FIG. 4. Dependence of retardance on laser parameters. (a) and (b) Retardance maps as functions of pulse energy and pulse number for Yb: KGW laser (top) and fiber laser (bottom) with pulse duration of 300 fs and repetition rates (RR = 500 kHz or 1 MHz). The scan speed for RR = 500 kHz and 1 MHz is 1 mm/s and 2 mm/s, respectively. The areas within white dotted lines show the type X modification region.

On the other hand, at RR = 1 MHz and $E_p$ = 500 nJ, type X voxels can be created by the fiber laser, however in a narrower - compared to the Yb:KGW laser - pulse energy range [Fig. 4b]. At a longer pulse duration of 500 fs and RR = 500 kHz, the type X modification was observed in the pulse energy range from 550 nJ to 1 µJ for the Yb:KGW laser and was not observed with fiber laser [Fig. S1a]. At RR = 1 MHz and $\tau$ = 500 fs, the parameter window for X-type formation with the fiber is narrower - similarly to the 300 fs long pulses - than that for Yb:KGW laser [Fig. S1b]. The difference between the characteristics of the fiber laser at RR = 500 kHz and RR = 1 MHz can be explained by the accumulation of third-order nonlinear effects in the fiber amplifier chain, for example, self-phase modulation, produced with greater efficiency by ultrashort light pulses with a higher intensity at RR = 500 kHz, leading to the formation of a subnanosecond pedestal.

To verify this hypothesis, we characterized the temporal contrast of pulses from two lasers by an autocorrelator having the maximum dynamic range of $10^7$. At a repetition rate of 500 kHz, the temporal contrast of pulses from the Yb:KGW laser was around $10^7$, i.e., produced pulses have no noticeable pedestal [Fig. 5a]. In contrary, at RR = 500 kHz, the temporal contrast of the fiber laser pulses was as low as $10^3$. That is at the pulse duration of 300 fs and the pedestal width of about 200 ps, the pedestal carries nearly 32% of the pulse energy. Correspondingly, at the pulse energy of 700 nJ, the peak powers for the femto- and subnanosecond components of the pulse are 1.6 MW and 1.1 kW, respectively. Thus, the actual intensity of focused femtosecond pulses produced by Yb:KGW and the fiber lasers are 10 TW/cm$^2$ and 6.8 TW/cm$^2$, respectively. The latter explains why the threshold pulse energy required for the material modification with the fiber laser is 1.5 times higher than that of Yb: KGW laser. Moreover, the parameters window for isotropic refractive index increase is also broader for high contrast pulses [Fig. S2].

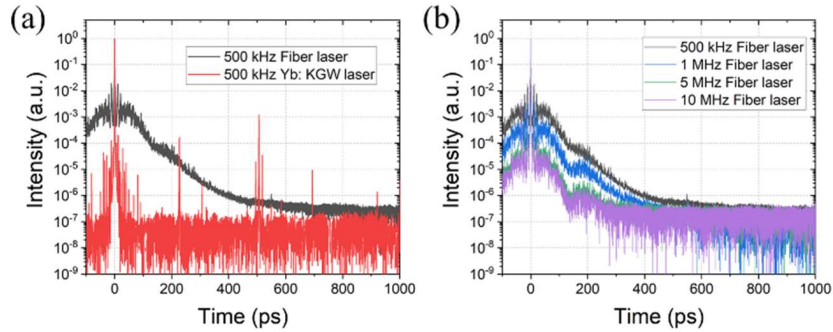

FIG. 5. Temporal contrast ratio of 300 fs pulses from Yb: KGW and fiber lasers over a time range of -100 to 1000 ps. (a) Comparison of the contrast ratio at a repetition rate of 500 kHz for both lasers. (b) Contrast ratio for fiber laser at various repetition rates.

We also compared the temporal contrast of the fiber laser pulses at repetition rates of 500 kHz, 1 MHz, 5 MHz and 10 MHz. One can observe from Fig. 5b that the higher the RR, the better the temporal contrast because increasing the repetition rate results in the suppressing self-phase modulation of the low peak power pulse in the fiber laser amplifier chain. However, even at RR = 10 MHz the achievable contrast of $10^5$ is still two orders lower than that observed for the Yb:KGW laser at RR = 500 kHz. The temporal contrast can be improved by using appropriate phase compensation techniques.

Lamella shaped birefringent (type 2m) modification, which is stronger and observed at a lower threshold for fiber laser may originate from the absorption of the subnanosecond pedestal that carries about third of the pulse energy. Since the lifetime of free electrons in silica glass is less than 1 ps,[28,29] one can expect that the existence of an alternative light absorption channel. We believe that the key role is played by self-trapped holes (STHs),[30] which have short lifetime at room temperature and low excitation energy (1-2 eV).[31,32] Created in the irradiated volume by femtosecond portion (68% in energy) of the fiber laser pulse, they are capable of absorbing a significant part of the pulse pedestal energy due to one- and two-photon absorption, promoting the formation of lamella-shaped structures rather than nanopores in the irradiated volume.

One may also suggest that the nonlinear current defined by the concentration gradient and light polarization, which reads $\boldsymbol{j_q} \propto (\boldsymbol{\nabla} n \boldsymbol{E})\boldsymbol{E}$

,[33,34] where *n* is concentration of charge carriers and **E** is the electric field of light beam, participates in the formation of lamella-shaped structures oriented perpendicular to the polarization and possibly plays a key role in nanograting formation. Indeed, this nonlinear current is an analog of the quadrupole nonlinear polarization, which is responsible, for example, for the generation of the second harmonic in an isotropic medium. Interestingly, this nonlinear current, determined by the polarization of the light beam, is also directed against the charge concentration gradient, which leads to its amplification. The latter can explain the creation of narrow lamella-like structures oriented perpendicular to the direction of light polarization. Moreover, the evanescent near-field component along the polarization direction created by lamella-like nanostructures propagates with a speed of light along the polarization and can interfere with the incident light field [35]. Such an interference results in the spatial modulation of the light intensity and concentration of light-induced defects, for example, self-trapped holes with a wavelength periodicity along the direction of polarization. Earlier, the mechanism of formation of structures with a periodicity of the wavelength was also discussed.[36]

In conclusion, we demonstrated that the pulse temporal contrast can be used for controlling ultrafast laser writing in silica glass. The low pulse contrast of $10^3$, which is characteristic of an amplified femtosecond fiber laser, results in a higher energy threshold for ultrafast laser writing compared to high-contrast $10^7$ pulses from a regeneratively amplified Yb: KGW laser system. Optically invisible birefringent modification with nanopore structure cannot be created with low-contrast pulses. Instead, lamella-shaped birefringent structures (type 2m) with larger birefringence are produced, which can be attributed to the absorption of low power subnanosecond pulse pedestal by transient defects with low excitation energy, such as self-trapped holes, generated by high-power femtosecond part of the pulse. Such lamella-like structures can be beneficial for applications in etching-assisted micromachining of silica glass.[4] Periodic structures with wavelength periodicity are observed, indicating the material modification mechanism, involving the interference of the light field of the writing laser beam with the evanescent near-field component created by the nanostructure and propagating along the polarization direction, an alternative to the interference of the incident and the inhomogeneity-scattered light waves.

See the **Supplementary Material** for the classification of laser-induced birefringent modifications and dependence of laser writing parameters on type X and type 1 modifications.

The authors acknowledge Mateusz Ibek from APE GmbH for lending us the high dynamic range autocorrelator. We acknowledge Clemens Hoenninger from Amplitude and Martynas Baskauskas from Light Conversion for useful discussions. This research was supported by European Research Council (ENIGMA, 789116); Microsoft (Project Silica); Horizon 2020 Marie Curie RISE (CHARTIST, 101007896); Academy of Finland (decision nos. 343393, 320166).

**AUTHOR DECLARATIONS**

**Conflict of Interest**

The authors have no conflicts to disclose.

# Supplementary Material

## Controlling ultrafast laser writing in silica glass by pulse temporal contrast

In our experimental conditions (1030 nm wavelength, 300 fs pulse duration, 200 kHz repetition rate, 700 nJ pulse energy, and 0.16 NA aspherical lens), randomly distributed flattened nanopores (type X modification) form at $20 < N_p < 150$. When the number of pulses exceeds 150 ($N_p > 150$), flattened nanopores are transformed into randomly distributed nanolamellae (type X-2 modification). Further increase of the number of pulses ($N_p > 300$) results in the formation of lamella like structures with $\lambda/n$ periodicity (type 2m modification). At even higher number of pulses, we arrive at well-known self-assembled nanogratings with subwavelength periodicity (type 2 modification). Another special birefringent modification is single nanolamella structures, in which an isotropic nanovoid is initially generated by microexplosion with a high-energy pulse and then reshaped to a nanolamella-like structure by near-field enhancement with lower energy pulses.

Table S1. Classification of laser-induced birefringent modifications.

| Classification | Morphology revealed by SEM | Visibility in optical image |
|---|---|---|
| Type X | Randomly distributed flattened nanopores | No |
| Type X-2 | Randomly distributed nanolamella | Yes, due to weak scattering |
| Type 2m | Lamella shaped structures with $\lambda/n$ period | Yes, with $\lambda/n$ periodicity |
| Type 2 | Self-assembled nanogratings with subwavelength period | Yes, due to stronger scattering |
| Type S | Single nanolamella-like structures | Yes, due to weak scattering |

To reveal the difference in the parameter windows for obtaining the type X modification, experiments were carried out with two lasers using different repetition rates, pulse numbers, and energies. The pulse duration was 500 fs. At repetition rate (RR) of 500 kHz, the anisotropic nanopores based type X modification was imprinted in the pulse energy range from 550 nJ to 1 µJ for the Yb: KGW laser, but it was not observed with the fiber laser (Fig. S1a). At RR of 1 MHz, type X modification was produced with the fiber laser with a smaller parameter window compared to the Yb: KGW laser (Fig. S1b).

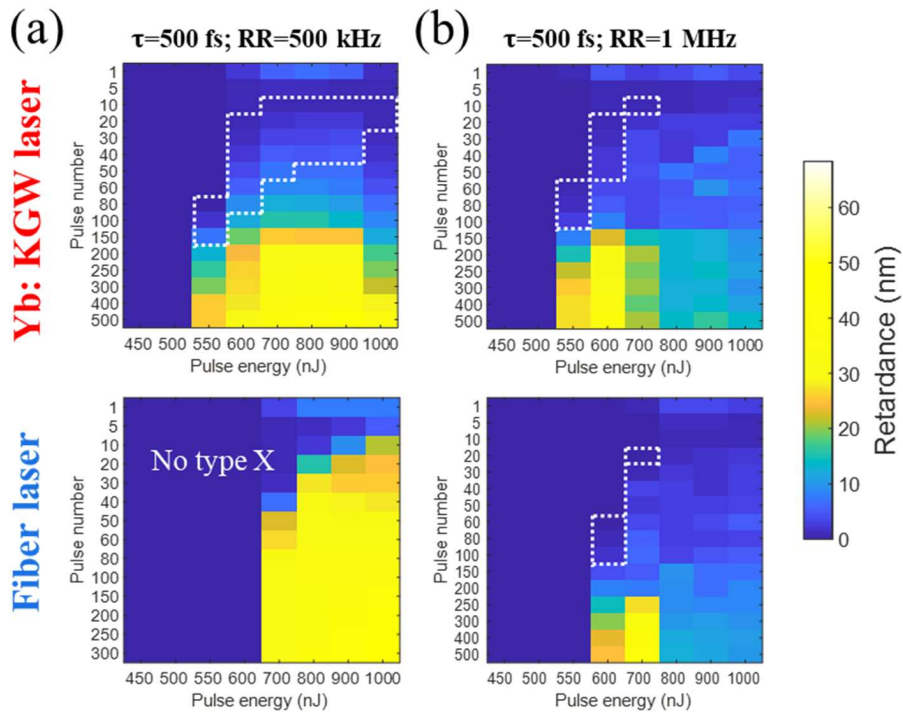

Fig. S1. Dependence of retardance on laser parameters at pulse duration of 500 fs. (a) and (b) Retardance maps as functions of pulse energy and pulse number for Yb: KGW laser (top) and fiber laser (bottom) with repetition rates (RR = 500 kHz or 1 MHz). The scan speed for RR = 500 kHz and 1 MHz is 1 mm/s and 2 mm/s, respectively. The areas within white dotted lines show the type X modification region.

We also performed quantitative comparison of Yb: KGW and fiber lasers in terms of their ability to write structures with the refractive index increase (type 1 modification) in silica glass. The experiments were carried out using a 0.4 NA lens, the pulse duration of 270 fs and the repetition rate of 100 kHz, pulse energies from 125 nJ to 300 nJ and scanning speeds ranging from 10 mm/s to 50 mm/s.

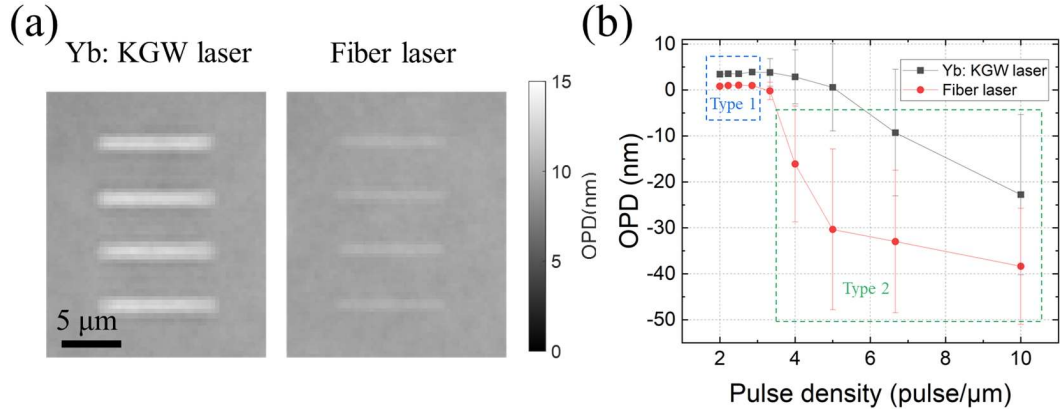

Fig. S2. Comparison of isotropic refractive index increase (type 1) written by the Yb: KGW and fiber lasers. (a) Phase image of type 1 modification written by the two lasers and their spatial optical path difference (OPD). The pulse energy was 175 nJ and the scanning speed was 45 mm/s, corresponding to 2.2 pulses per µm. (b) OPD as a function of pulse density for both lasers. Processing conditions: 1030 nm wavelength, 270 fs pulse duration, 100 kHz repetition rate, focusing via 0.4 NA lens.

Figure S2 shows that, as for the type X modification, the parameter window for the type 1 modification with the Yb:KGW laser is wider than with the fiber laser. Specifically, the Yb:KGW laser was able to produce type 1 modification at pulse energies ranging from 125 nJ to 200 nJ, and scanning speeds ranging from 15 mm/s to 50 mm/s, while the fiber laser was only able to induce type 1 modification at a pulse energy of 175 nJ and a scanning speed from 35 mm/s to 50 mm/s. In addition, the optical path difference (OPD) of type 1 line structures written by the Yb: KGW laser was greater than that of the fiber laser at the same laser writing parameters (Fig. S2a). For example, the measured OPD was 3.5 (0.9) nm for type 1 line written by the Yb: KGW (fiber) laser at the pulse density of 2.2 pulses/µm (Fig. S2b). Moreover, when writing with the Yb: KGW laser with a pulse density of more than 6 pulses/µm, we observed the formation of nanograting type modification (type 2) with a negative average refractive index change and corresponding OPD. In contrast, modifications with more negative OPD were imprinted with fiber laser pulses with the same writing parameters, implying stronger type 2 modifications.